\documentclass[useAMS,usenatbib]{mn2e}
\usepackage{graphicx}
\DeclareGraphicsExtensions{.pdf}
\usepackage{amsfonts,amsmath,amssymb}
\usepackage[colorlinks=true,linkcolor=red, citecolor= blue, urlcolor=cyan]{hyperref}
\usepackage{upgreek}

\newcommand {\aplt} {\ {\raise-.5ex\hbox{$\buildrel<\over\sim$}}\ } 
\newcommand{\be}{\begin{equation}}
\newcommand{\ee}{\end{equation}}

\newfont{\gwpfont}{cmssq8 scaled 1000}

\newcommand{\mincir}{\raise
  -2.truept\hbox{\rlap{\hbox{$\sim$}}\raise5.truept \hbox{$<$}\ }}
\newcommand{\magcir}{\raise
  -2.truept\hbox{\rlap{\hbox{$\sim$}}\raise5.truept \hbox{$>$}\ }}
\newcommand{\siml}{\raise
  -2.truept\hbox{\rlap{\hbox{$\sim$}}\raise5.truept \hbox{$<$}\ }}
\newcommand{\simg}{\raise
  -2.truept\hbox{\rlap{\hbox{$\sim$}}\raise5.truept \hbox{$>$}\ }}

\newcommand{\XMM}{XMM-{\it Newton}}

\defcitealias{2019bCapasso}{C19}

\begin{document}

\title[Calibration of CODEX Luminosity-Mass relation]{Mass Calibration of the CODEX Cluster Sample using SPIDERS Spectroscopy - II. The X-ray Luminosity-Mass Relation}

\newcommand{\Munich}{$^{1}$}
\newcommand{\ExcellenceCluster}{$^{2}$}
\newcommand{\INAFTrieste}{$^{3}$}
\newcommand{\IFPU}{$^{4}$}
\newcommand{\Stockholm}{$^{5}$}
\newcommand{\MPE}{$^{6}$}
\newcommand{\AstroTrieste}{$^{7}$}
\newcommand{\IRAP}{$^{8}$} 
\newcommand{\Helsinki}{$^{9}$} 
\newcommand{\Liverpool}{$^{10}$}

\author[Capasso et al.] {R.~Capasso\thanks{raffaella.capasso@fysik.su.se}\Munich$^,$\ExcellenceCluster$^,$\INAFTrieste$^,$\IFPU$^,$\Stockholm,
J.~J.~Mohr\Munich$^,$\ExcellenceCluster$^,$\MPE,
A.~Saro\INAFTrieste$^,$\IFPU$^,$\AstroTrieste,
A.~Biviano\INAFTrieste,
N.~Clerc\MPE$^,$\IRAP,
\newauthor
A.~Finoguenov\MPE$^,$\Helsinki,
M.~Klein\Munich$^,$\MPE,
S.~Grandis\Munich$^,$\ExcellenceCluster,
C.~Collins\Liverpool,
S.~Damsted\Helsinki,
\newauthor
C.~Kirkpatrick\Helsinki,
A.~Kukkola\Helsinki
\\
\\
%\date{\textit{Affilitations are listed at the end of the paper}}
\Munich Faculty of Physics, Ludwig-Maximilians-Universit\"{a}t, Scheinerstr.\ 1, 81679 Munich, Germany \\
\ExcellenceCluster Excellence Cluster Universe, Boltzmannstr.\ 2, 85748 Garching, Germany \\
\INAFTrieste INAF-Osservatorio Astronomico di Trieste, via G. B. Tiepolo 11, I-34143 Trieste, Italy \\
\IFPU IFPU - Institute for Fundamental Physics of the Universe, Via Beirut 2, 34014 Trieste, Italy \\
\Stockholm The Oskar Klein Centre, Department of Physics, Stockholm University, Albanova University Center, SE 106 91 Stockholm, Sweden \\
\MPE  Max Planck Institute for Extraterrestrial Physics, Giessenbachstr.\ 85748 Garching, Germany \\
\AstroTrieste Astronomy Unit, Department of Physics, University of Trieste, via Tiepolo 11, I-34131 Trieste, Italy \\
\IRAP IRAP, Université de Toulouse, CNRS, UPS, CNES, 31400 Toulouse, France \\
\Helsinki Department of Physics, University of Helsinki, Gustaf H\"{a}llstr\"{o}min katu 2a, FI-00014 Helsinki, Finland \\
\Liverpool Astrophysics Research Institute, Liverpool John Moores University, IC2, Liverpool Science Park, 146 Brownlow Hill, Liverpool, L3 5RF, UK
}

\pubyear{2019}

\date{\today}

\maketitle               
%\saythanks      

\label{firstpage}

\begin{abstract} 
We perform the calibration of the X-ray luminosity--mass scaling relation on a sample of 344 CODEX clusters with $z<0.66$ using the dynamics of their member galaxies. Spectroscopic follow-up measurements have been obtained from the SPIDERS survey, leading to a sample of 6,658 red member galaxies. We use the Jeans equation to calculate halo masses, assuming an NFW mass profile and analyzing a broad range of anisotropy profiles. 
With a scaling relation of the form $L_{\rm{X}} \propto \text{A}_{\rm{X}}M_{\text{200c}}^{\text{B}_{\rm{X}}} E(z)^2 (1+z)^{\gamma_{\rm{X}}}$, 
we find best fit parameters $ \text{A}_{\rm{X}}=0.62^{+0.05}_{-0.06} (\pm0.06)\times10^{44}\,\mathrm{erg\,s^{-1}}$, $\text{B}_{\rm{X}}=2.35^{+0.21}_{-0.18}(\pm0.09)$, $\gamma_{\rm{X}}=-2.77^{+1.06}_{-1.05}(\pm0.79)$, where we include systematic uncertainties in parentheses and for a pivot mass and redshift of $3\times10^{14}M_\odot$ and 0.16, respectively.  We compare our constraints with previous results, and we combine our sample with the SPT SZE--selected cluster subsample observed with \XMM\, extending the validity of our results to a wider range of redshifts and cluster masses.
\end{abstract}


%%%%%%%%%%%%%%%%%%%%%%%%%%%%%%%%%%%%%%
%%%% %%%                       Introduction                               %%%%%%%%
%%%%%%%%%%%%%%%%%%%%%%%%%%%%%%%%%%%%%%

\section{Introduction}  
\label{sec:introduction}

Accurate mass estimates of galaxy clusters are of fundamental importance for both cosmological and astrophysical studies. Observational knowledge of the mass distribution of the dark and baryonic matter in clusters provides insights into their formation and evolution \citep[see, e.g.][]{2001Springel, 2004bGao, 2015Popesso, 2019Pratt}. On the other hand, number counts of galaxy clusters, sensitive to the amplitude of matter fluctuations, can provide constraints on various cosmological parameters in a way complementary to other cosmological probes \citep[e.g.,][]{1993White,2001Haiman,2015Mantz,2018Bocquet}. 
Studies of the link between the observable features of haloes and the underlying matter distribution are thus essential. \\

An efficient use of clusters as cosmological probes requires a low-scatter mass proxy to relate theoretical predictions to observations \citep{2005Lima, 2011Allen}. 
To infer the mass of a sample of galaxy clusters we have to be able to characterize a number of biases, depending on the intrinsic covariance of the cluster observables, measurement uncertainties and selection effects \citep[e.g.][]{2007Pacaud, 2010Mantz, 2016deHaan}.
 The combination of limited surveyed volume and source selection thresholds produce the well known Malmquist bias \citep{1920Malmquist}, truncating the scattered distributions of sources in the space of observables. As a consequence, luminosity or flux selected samples are typically biased towards low masses where the selection is returning only a fraction of the underlying cluster sample. This effect is enhanced by the so-called Eddington bias \citep{1913Eddington}.  Because the number density of halos is a steeply falling function of their mass  \citep[e.g.][]{2008Tinker,2016Bocquet}, the presence of scatter in the relationship between the selection observable (i.e., flux or luminosity) and mass will cause low-mass clusters to preferentially up-scatter, leading to a bias in the mass associated with the observable \citep{2011Mortonson}.  An accurate calibration of cluster scaling relations requires control over these biases. \\

Many different mass proxies have been used over the years, including thermal Sunyaev--Zeldovich effect (SZE) measurements \citep{2009Staniszewski, 2014PlanckXX,2013Hasselfield}, weak gravitational lensing features \citep{2009Corless, 2011Becker, 2018Dietrich}, cluster velocity dispersions \citep{2006Biviano, 2013Saro, 2019aCapasso}, X-ray luminosity and temperature \citep{2009Vikhlinin, 2010Mantz, 2016Andreon}, and $Y_{X}$ parameter, i.e. the product of the X-ray temperature and gas mass \citep{2014Maughan, 2016Mantz, 2018Mantz, 2018Truong}.  A combination of multiple, independent mass proxies help mitigate systematic uncertainties \citep{2015Bocquet,  2018McClintock, 2018Baxter, 2018Farahi, 2018Bocquet}.
In a companion paper \citep[][hereafter C19]{2019bCapasso} we performed the dynamical mass calibration exploiting the optical richness of a sample of 428 CODEX \citep[COnstrain Dark Energy with X-ray clusters;][]{2019Finoguenov} clusters, constraining the amplitude of the $\lambda$--mass relation with a $\sim$12\% accuracy.

Following \citetalias{2019bCapasso}, we calibrate the X-ray luminosity--mass--redshift scaling relation by exploiting the information residing in the observed projected phase space (distribution in line of sight velocities and projected radius) of the cluster member galaxies.  We use a modification of the MAMPOSSt  technique \citep[Modeling Anisotropy and Mass Profiles of Observed Spherical Systems;][]{2013MAMPOSSt}, based on the Jeans equation \citep{1987Binney}, to simultaneously determine the dynamical cluster masses and the parameters of the scaling relation. The MAMPOSSt code has been successfully used to investigate the internal dynamics of clusters, determining their masses and velocity anisotropy profiles \citep[e.g.][]{2013Biviano,2017Biviano, 2014Munari, 2019aCapasso}. 

We perform this analysis on the CODEX cluster catalog, which consists of ROSAT All-Sky Survey (RASS) X-ray cluster candidates having optical counterparts in SDSS imaging data identified using the RedMaPPer algorithm \citep[the red-sequence Matched-filter Probabilistic Percolation algorithm,][]{2014Rykoff}. A subset of this sample has been spectroscopically studied within the SPectroscopic IDentification of eRosita Sources (SPIDERS) survey \citep{2016Clerc}. 
The analysis we carry out focuses on a sample of 344 CODEX clusters with a corresponding sample of $\sim$6600 red member galaxies with measured redshifts.  The clusters span the redshift range $0.03 \leq z_{\text{c}}  \leq 0.66$, with richnesses $ 20 \leq \lambda \leq 230$ and rest-frame [0.1-2.4]~keV luminosities $ 4.5\times10^{42} \leq  L_{\rm{X}}/(\rm{erg\,s}^{-1})  \leq 3.2\times10^{45} $. 

The paper is organized as follows. In Section~\ref{sec:theory} we detail the theoretical framework. In Section~\ref{sec:data} we present the data set used in this analysis and the selection criteria. The likelihood model used to constrain the $L_{\rm{X}}$--mass--redshift scaling relation is described in Section~\ref{sec:results}, followed by the outcome of our calibration, and a discussion of a range of systematic uncertainties. We present our conclusions in Section~\ref{sec:conclusions}.

Throughout this paper we assume a flat $\Lambda$CDM cosmology with a Hubble constant $H_{0} = 70 \,  \text{km} \, \text{s}^{-1} \,  \text{Mpc}^{-1}$, and a matter density parameter $\Omega_{\text{M}} = 0.3$. Cluster masses ($M_{\text{200c}}$) are defined within $r_\mathrm{200c}$, the radius of the sphere inside which the cluster overdensity is 200 times the critical density of the Universe at the cluster redshift. We refer to $r_\mathrm{200c}$ as the virial radius. All quoted uncertainties are equivalent to Gaussian $1\sigma$  confidence regions, unless otherwise stated.

%%%%%%%%%%%%%%%%%%%%%%%%%%%%%%%%%%%%%%
%%%% %%%                            Theory                                  %%%%%%%%
%%%%%%%%%%%%%%%%%%%%%%%%%%%%%%%%%%%%%%

\section{Theoretical Framework}
\label{sec:theory}

We use dynamical constraints on a large ensemble of clusters to constrain the underlying halo masses, thereby enabling measurement of the luminosity--mass-redshift relation.  To do this, we perform a dynamical analysis based on the application of the Jeans equation to spherical systems \citep{1987Binney}. The Jeans equation allows us to define the mass distribution $M(r)$ of a cluster as
\be                                                             
\label{eq:jeans}                                                              
\frac{GM(<r)}{r} = - {\sigma_{r}^{2}} \left( \frac{d \ln \nu}{ d \ln r } + \frac{d \ln {\sigma_{r}^{2}}} { d \ln r} + 2 
\beta \right) ,
\ee
with $\nu(r)$ being the number density profile of the tracer galaxy population, $\sigma_{r}(r)$ the radially dependent component of the velocity dispersion along the spherical coordinate $r$, $M(<r)$ the enclosed mass within radius $r$, $G$ Newton's constant, $\beta(r) \equiv 1 - (\sigma_{\theta}^{2} / \sigma_{r}^{2})$  the radially dependent velocity dispersion anisotropy, and $\sigma_{\theta}$, one of the two (assumed identical) tangential components of the velocity dispersion.

Equation~\ref{eq:jeans} can thus be used to estimate the mass distribution of a spherical system. However, the only observables we can directly obtain are projected quantities: the surface density profile of the galaxy distribution, the rest-frame LOS velocities and the radial separation of each galaxy from the cluster center. 
Because of projection effects, the determination of the mass distribution of a galaxy cluster is degenerate with the determination of the velocity anisotropy profile  \citep[e.g.][]{1987Merritt}.

In this work, we address this problem by applying the Modeling Anisotropy and Mass Profiles of Observed Spherical Systems algorithm \citep[hereafter MAMPOSSt; for full details please refer to][]{2013MAMPOSSt}. This method consists in determining the mass and anisotropy profiles of a cluster in parametrized form by performing a likelihood exploration of the distribution of the cluster galaxies in projected phase space, comparing it to the theoretical distribution predicted from the Jeans equation for these models. This method thus requires adopting parametrized models for the number density, mass, and velocity anisotropy profiles $\nu(r)$, $M(r)$, $\beta(r)$. 

As addressed in Section~\ref{sec:number_density}, because our spectroscopic dataset is likely to suffer from radially dependent incompleteness, we adopt the number density profile derived from a study of red sequence galaxies in SZE selected clusters \citep{2017Hennig}.  

Regarding our choice of the mass and velocity anisotropy profiles, we follow our previous work \citetalias{2019bCapasso}. We refer to that study for a more detailed description. In the next section we summarize the main features.

\subsection{Mass and anisotropy profiles}
\label{sec:profiles}

Driven by both numerical studies of structure formation and observational results, we adopt the mass model introduced by \citet[][NFW]{1996NFW}, which is fully described by two parameters: the virial radius $r_{200}$, and the scale radius $r_{\rm{s}}$, which is the radius at which the logarithmic slope of the density profile is $-2$. Numerous observational studies have indeed found the mass distributions of clusters to be well described by this model \citep{1997Carlberg, vanderMarel2000, 2003Biviano, Katgert2004, 2011Umetsu}. 

On the other hand, due to the lack of published studies providing strong predictions for the radial form of the velocity anisotropy profile $\beta(r)$, we consider five models that have been used in previous MAMPOSSt analyses, described also in \citet{2019aCapasso}:  (1) constant anisotropy model (C), (2) Tiret anisotropy profile \citep[][T]{2007Tiret}, (3) \citet{2005Mamon} profile (M{\L}), (4) Osipkov-Merritt  anisotropy profile \citep[][OM]{1979Osipkov,1985Merritt}, and (5) a model with anisotropy of opposite sign at the center and at large radii (O). 

Therefore, given a mass for each cluster we predict the projected phase space distribution of the observed dynamical dataset by running MAMPOSSt with 3 free parameters: the virial radius $r_\mathrm{200c}$, the scale radius $r_\mathrm{s}$ of the mass distribution, and a velocity anisotropy parameter $\theta_{\beta}$. The latter represents the usual $\beta=1-(\sigma_{\theta}^{2}/\sigma_{r}^{2})$ for the first three models (C, T, O), while for the $\text{M{\L}}$ and OM models it defines a characteristic radius $\theta_\beta=r_\beta$.  

\subsection{Bayesian model averaging}
\label{sec:Bayes}

As described above, we employ five velocity anisotropy models when estimating the projected phase space distribution of member galaxies for each cluster.  Because we cannot strongly reject any of the models, we combine the results obtained from each anisotropy model $\beta(r)$ by merging their constraints, exploiting the Bayesian model averaging technique \citep[see C18,][for more details]{2019aCapasso}. In a nutshell, this method consists in assigning a weight to each model, according to how well the model fits the data.  This weight is represented by the so-called Bayes factor \citep[see][and references therein]{Hoeting99bayesianmodel}. 

Considering the 5 anisotropy models $M_{1}$, ..., $M_{5}$, the Bayes factor $B_{j}$ of each model $j$ is defined as the marginalized likelihood of the model $\mathcal{L}(D \,| M_{j})$, also known as evidence, normalized by the likelihood of the most probable model.  Specifically, 
\be
\label{eq:bayesfactor}
B_{j} = {\mathcal{L}(D \,| M_{j}) \over \mathcal{L}(D \,| M_{\text{max}})}  ,
\ee
where $M_{\text{max}}$ indicates the model with the highest marginalized likelihood, $\mathcal{L}(D \,| M_{j}) = \int { \mathcal{L}(D\,| \theta_{j}, M_{j}) P(\theta_{j}\,| M_{j}) \,\, d \theta_{j} }$, $\mathcal{L}(D\,| \theta_{j}, M_{j})$ is the likelihood of the data $D$ given the model parameters $ \theta_{j}$, and $P(\theta_{j}\,| M_{j})$ is the prior. 

The average posterior distribution of the fitted scaling relation parameters is then given by the weighted average of the posterior distributions of each model, with the Bayes factor as weight. This Bayesian model averaging is performed by means of the multimodal nested sampling algorithm MultiNest \citep{2008Feroz, 2009Feroz, 2013Feroz}, providing us with the evidence for each model.

%%%%%%%%%%%%%%%%%%%%%%%%%%%%%%%%%%%%%%
%%%% %%%                             Data                                    %%%%%%%%
%%%%%%%%%%%%%%%%%%%%%%%%%%%%%%%%%%%%%%

\section{Data}
\label{sec:data}

We perform our analysis on a subset of CODEX galaxy clusters observed within the SPIDERS survey \citep{2016Clerc}, which provides us with the spectroscopic galaxy sample. The CODEX sample is based on ROSAT All-Sky Survey \citep[RASS, see][]{1999Voges} selected clusters, cross--matched with nearby optically selected systems identified using the redMaPPer \citep[the red-sequence Matched-filter Probabilistic Percolation,][]{2014Rykoff} algorithm applied to the Sloan Digital Sky Survey IV \citep[SDSS-IV, see][]{2016Dawson, 2017Blanton} optical imaging data. A full description of the dataset construction and features are described in \citetalias{2019bCapasso}. In the following section we summarize the main elements of the dataset.

\subsection{The CODEX sample}
\label{sec:CODEX}

The CODEX cluster sample combines ROSAT X-ray cluster candidates with optical selected cluster candidates identified using redMaPPer. First of all, RASS data are searched for all X-ray sources with detection significance S/N$>4$. Then, redMaPPer is run on the SDSS imaging data around each of these sources. RedMaPPer is an optical cluster-finding algorithm based on the red sequence technique, built around the richness estimator of \citet{2012Rykoff}. This step thus allows the identification of candidate clusters with a red-sequence, constituting a collection of passive galaxies at a common redshift. The redMaPPer algorithm provides an estimate for the cluster photometric redshift, an estimation of the optical richness and an optical cluster center.  In cases of multiple optical counterparts meeting these criteria, the counterpart having the highest richness is assigned to the RASS X-ray source.  
The updated optical cluster position allows the identification of a revised red-sequence, providing the final estimate of the cluster photometric redshift and richness. 
Finally, RASS count-rates provide an estimate for the X-ray properties of the clusters. Assuming a model for the X-ray spectral emissivity, imposing a minimal S/N threshold of 1.6 to have optimized apertures, we calculate the aperture-corrected cluster flux $f_{X}$. 
From the flux, we iteratively recover the X-ray luminosities $L_{X}$ in the rest-frame 0.1-2.4 keV band. Starting from an initial luminosity given by the flux, we obtain an initial guess on the cluster mass and temperature using the XXL $M - T$ \citep{2016Lieu} and $L_{X} -T$ \citep{2016Giles} scaling relations. From the cluster mass we compute $R_{500}$ using the concentration-mass relation of \citet{2014Dutton}, which leads to an updated flux extraction aperture and a new estimate of the luminosity. The iteration continues until convergence is reached. For a full description of this procedure, we refer the reader to \citet{2019Finoguenov}.

The final CODEX sample is then characterized by X-ray detected clusters, with estimated redshift, optical richness, optical cluster center, and X-ray luminosity.   Follow-up observations obtained with the SPIDERS survey, described below, finally provide us with spectroscopic redshift redshift measurements of cluster member galaxies.

\subsection{The SPIDERS spectroscopic sample}
\label{sec:spiders}

The SPIDERS survey is designed to obtain homogeneous and complete spectroscopic follow-up of X-ray extragalactic sources lying within the SDSS-IV imaging footprint, with the aim of confirming galaxy cluster candidates and of assigning a precise redshift measurement. In particular, this survey was conceived to obtain follow-up observations of X-ray extended sources extracted from the all sky X-ray eROSITA survey \citep[extended ROentgen Survey with an Imaging Telescope Array;][]{2010Predehl, 2012Merloni}. However, prior to the launch of eROSITA, the bulk of the SPIDERS program galaxy clusters is made of those identified in the shallower RASS and sparser \XMM\ data. 
At the time this paper is being written, the observations of these clusters have already been completed. No further galaxy spectroscopic redshifts will be assigned to them during the final stages of the SDSS-IV program.

The target selection is performed so as to optimize the number of spectroscopically confirmed clusters. As a first step, the redMaPPer membership probability is used to assign priorities to potential targets, ranking galaxies within each cluster. The pool of targets along with the priority flag is then submitted to the eBOSS tiling algorithm. The eBOSS spectroscopic pipeline is then employed to produce the final data reduction and spectral classification. 

For each cluster, an automatic procedure assigns membership of red-sequence galaxies with measured redshifts.  This is performed through an iterative clipping procedure. Members with rest-frame velocities (relative to the first guess cluster redshift) greater than 5000~km/s are rejected. The remaining potential members are used to estimate the velocity dispersion of the cluster. A $3\sigma$ clipping is then applied, rejecting objects lying further away than 3 times the velocity dispersion from the mean velocity. 

In the course of this iterative procedure, a few problematic cases typically occur.  For example, fewer than 3 members are sometimes assigned to a cluster,  and sometimes the initial 5000~km/s clipping rejects all members. In such cases the problematic cluster is flagged and visually inspected by independent inspectors. This final validation may lead to the inclusion or removal of members, as well as the identification of other structures lying along line-of-sight of the cluster. Final cluster redshift estimates are based on the bi-weight average \citep{1990Beers} of all galaxies selected as cluster members, if at least 3 members are assigned to the cluster. The cluster redshift statistical uncertainty is typically $\Delta_{z}/(1 + z) \lesssim 10^{-3}$.

Finally, the updated cluster spectroscopic redshifts are used to update the measurement of X-ray cluster properties. Assuming the standard flat $\Lambda$CDM cosmological model (Hubble constant $H_{0} = 70 \,  \text{km} \, \text{s}^{-1} \,  \text{Mpc}^{-1}$, and matter density parameter $\Omega_{\text{M}} = 0.3$, ROSAT fluxes are converted into rest-frame [0.1-2.4]~keV luminosities. The typical measurement uncertainty on the luminosities is $\approx$ 35\%, as computed from the Poissonian fluctuation in the associated ROSAT X-ray photons \citep[see][]{2015Mirkazemi}.

\subsection{Final spectroscopic cluster member sample}
\label{sec:finalmemberselection}

Before proceeding with our analysis, we apply some additional cuts to the SPIDERS spectroscopic sample.  First of all, to avoid systems that are clearly in a merging stage, we only use clusters which do not have any other component along the line of sight. 
As we are carrying out a Jeans analysis, based on the assumption of dynamical equilibrium, we restrict our analysis to the cluster virial region ($R  \leq r_\mathrm{200c}$). 
Moreover, we exclude the very central cluster region ($R \leq 50 \text{kpc}$), to avoid the inclusion of the central BCG, in which merger and dissipation processes could be ongoing, and to account for the positional uncertainties of cluster centers.
Our final spectroscopic dataset consists of 705 galaxy clusters, for a total of $\approx 11,400$ candidate cluster members, with a median redshift $z = 0.21$ and spanning an X-ray luminosity range $ 4.5\times10^{42} \leq  L_{\rm{X}}/(\rm{erg\,s}^{-1})  \leq 3.2\times10^{45} $, and a richness range $20 \leq \lambda \leq 230 $. 

We take the SPIDERS validated redshifts and redMaPPer positions to calculate the observable needed for our analysis:  the galaxy projected clustercentric distance R and the rest-frame line of sight (LOS) velocity $\mathrm{v}_\text{rf}$. 
Rest-frame velocities are then obtained as $\mathrm{v}_{\text{rf}} \equiv c(z_\text{gal} - z_\text{c})/(1+z_{\text{c}})$.

\begin{figure}
\centering 
\includegraphics[scale=0.35]{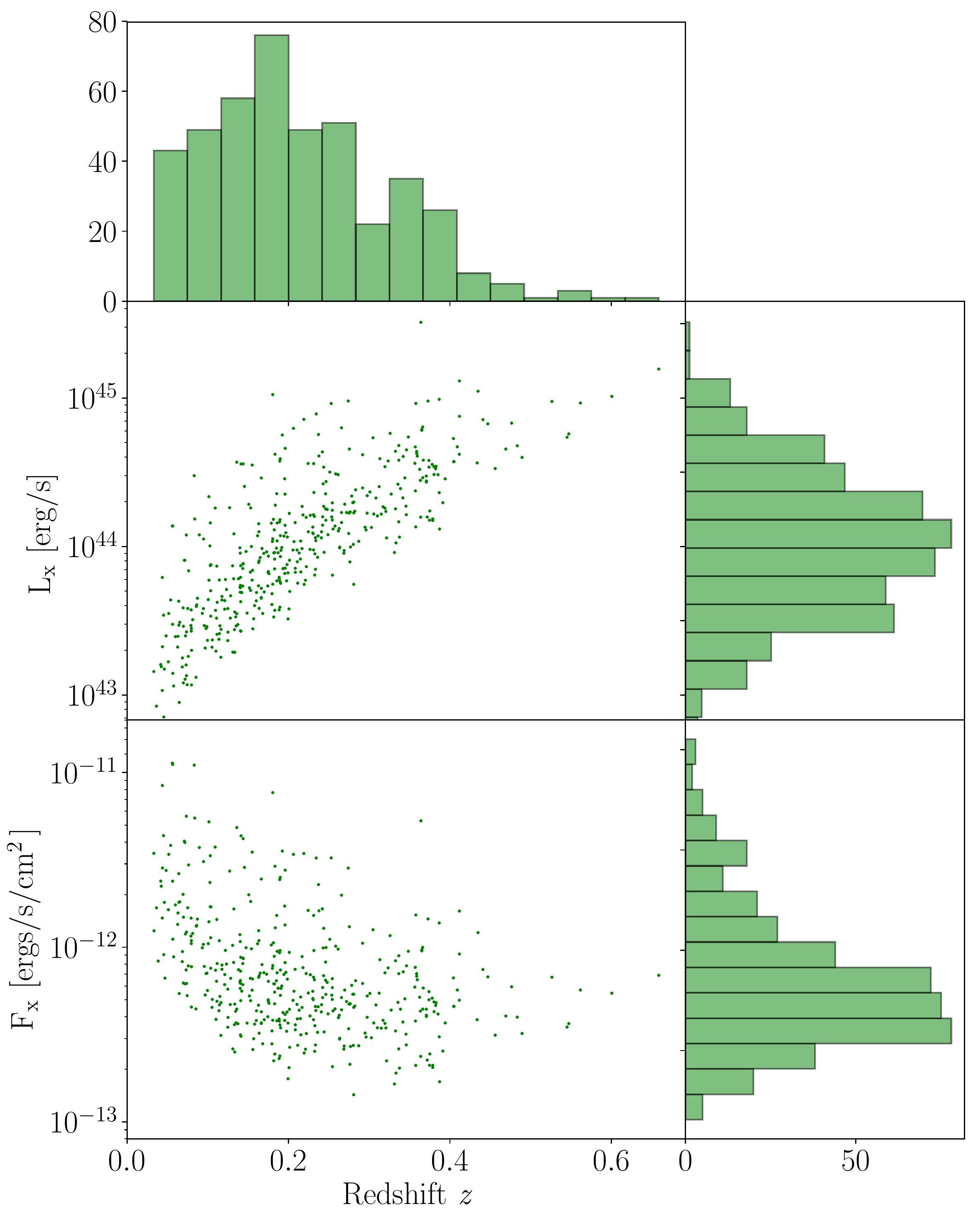}
\caption {Distribution of X-ray luminosity $L_{\rm{X}}$, flux $F_{\rm{X}}$, and cluster redshift $z_{\rm{c}}$ of the final cluster sample.}
\label{fig:histo}
\end{figure}

\subsection{Interloper rejection}
\label{sec:interloperrejection}

As described in Section~\ref{sec:spiders}, the SPIDERS automated procedure assesses membership for each galaxy in each cluster and subsequent visual inspection refines membership manually. However, interloper galaxies could still be present. These are galaxies that in projection are inside the cluster virial region, but do not actually lie inside it. We identify these objects by means of the  ``Clean" method \citep{2013MAMPOSSt}, based on the comparison between the location of the galaxies in the projected phase space and the expected maximal line of sight velocity at each projected radius. 
As we do not have enough spectroscopic redshifts to perform this method accurately for each individual cluster, we divide our sample into 15 equally spaced $\lambda$ bins, building a composite cluster in each bin. The composite clusters are built by stacking in metric radius [Mpc], without applying any scaling in velocity. We then perform the interloper rejection in each of them separately. 

The cleaning is performed in several steps. For each composite cluster, the LOS velocity dispersion $\sigma_{\text{LOS}}$ is used to estimate the cluster mass $M(r)$, using a scaling relation calibrated using numerical simulations \citep[e.g.,][]{2013Saro}, and assuming an NFW mass profile with concentration sampled from the mass--concentration relation. Then, assuming the M${\L}$ velocity anisotropy profile model, and given the cluster $M(r)$, an LOS velocity dispersion profile $\sigma_{\text{LOS}}(R)$ is calculated. Finally, galaxies with  $| \mathrm{v}_\text{rf} |  > 2.7 \sigma_{\text{LOS}}$ at any clustercentric distance are iteratively rejected \citep[see][]{Mamon2010, 2013MAMPOSSt}.

After the removal of interlopers, our spectroscopic sample consists of 703 clusters and 9,121 red galaxies. We apply a further cut on this dataset: we only keep systems that have at least 10 spectroscopic members, $N_{\text{mem}}\ge10$. This decision is driven by our concern that good constraints on the cluster masses and scaling relation parameters could not be obtained from clusters having very small numbers of spectroscopic members. We explore the impact of this cut in Section~\ref{sec:nmem}. 
After this cut, we are left with 428 clusters and 7807 red galaxies, with a median redshift, richness, and luminosity of $z = 0.16$, $\lambda$=41, and $\text{L}_{\text{X}}=  9.2 \times 10^{43}$erg\,s$^{-1}$, respectively. Fig.~\ref{fig:histo} shows the distributions of cluster redshift, X-ray luminosity and flux of the final sample.

We note that, even after this cleaning procedure, there is still a degree of contamination by interlopers. In general, galaxies lying outside the virial radius tend to have smaller peculiar velocities than those inside $R_{200}$.  Galaxies close to the cluster turn-around radius will have negligible peculiar velocities, and will not be identified as interlopers by the method adopted here.  An analysis of cosmological $N$-body simulations carried out by \citet{2013Saro} shows that, when passive galaxies are selected, this contamination is characteristically $\sim$20\% for massive clusters ($M_\mathrm{200c}\geq10^{14}M_\odot$), increasing with decreasing cluster mass.  
Another analysis carried out by \citet{Mamon2010} on hydrodynamical cosmological simulations shows that the distribution of interlopers in projected phase space is nearly universal, presenting only small trends with cluster mass. They find that, even after applying the iterative $2.7 \sigma_{\text{LOS}}$ velocity cut, the fraction of interlopers is still  23 $\pm$ 1\% of all DM particles with projected radii within the virial radius, and over 60\% between 0.8 and 1 virial radius. 

\subsection{Removing CODEX catalog contamination}
\label{subsec:contamination}

\begin{figure}
\centering { 
   \includegraphics[scale=0.55]{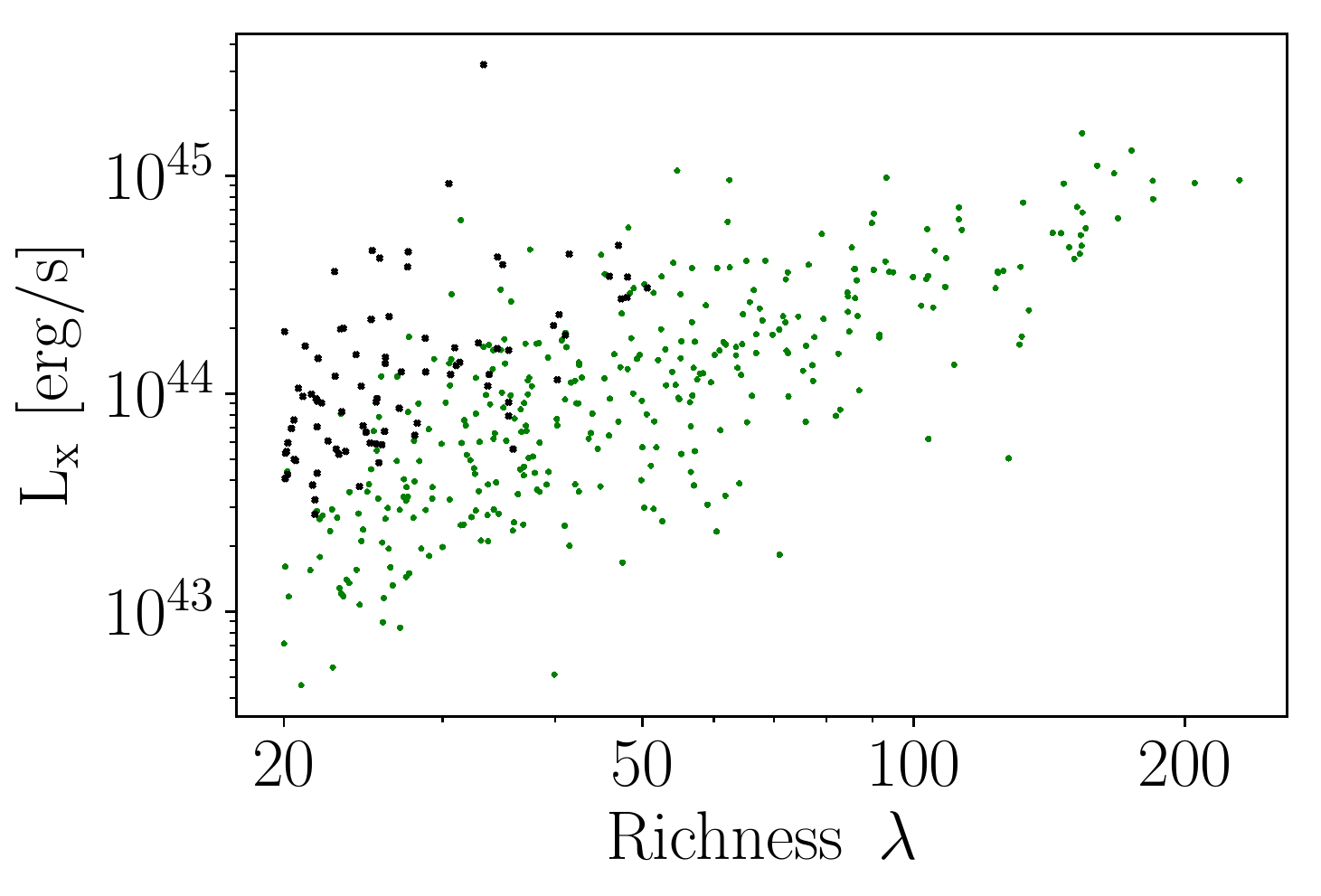} 
      }
\caption{Distribution of X-ray luminosity and richness for our cluster sample. Black crosses represent contaminated sources, which we exclude from our analysis. We imposed the cut  $f_{\rm{cont,m}}<0.05$, resulting in a catalogue with a 5\% contamination fraction.}
\label{fig:contamination}
\end{figure}

When cross-matching X-ray selected candidates from RASS with optical systems from redMaPPer or other similar techniques, one must be careful to account for the contamination of the resulting cluster catalog by random superpositions of physically unassociated X-ray and optical systems along the line of sight \citep{2018Klein}. For the RASS imaging, where there is generally no extent information for the faint CODEX sources, the contamination is driven by random superpositions between the faint X-ray sources ($\sim90$\% are AGN or stars) and the ubiquitous red sequence optical candidate clusters identified by redMaPPer. 

To exclude chance superpositions, we employ the method described in \citet{2019Klein}. This decontamination method consists of evaluating, for a cluster candidate at redshift $z$ and richness $\lambda$, the probability distribution of richness at that redshift for detected X-ray sources and that along random line of sights. Namely, we use the estimator $f_{\rm{cont}}$, which is defined as the ratio of the integral over the two distributions, above the observed $\lambda$ of the candidate \citep[see Fig.~6 and Eq.~10 in][]{2019Klein}. In particular, for a given richness and redshift, we adopt the value of $f_{\rm{cont, m}}$ from \citet{2019Klein}, which uses the distribution of observed richness together with the weighted mean of random richness distributions. We take into account differences in richness between the two RASS-based X-ray catalogues, assuming they have consistent selection and contamination properties.

We perform a cut at $f_{\rm{cont,m}}<0.05$, producing a sample with a 5\% contamination fraction, which is independent of redshift. After the cut, our final sample consists of 344 galaxy clusters with a total of 6,658 cluster members, characterized by a median redshift $z = 0.16$ and a median X-ray luminosity $ L_{\rm{X}} = 9\times10^{43} \,  \rm{erg\,s}^{-1} $. 
Figure~\ref{fig:contamination} shows the distribution of our sample as a function of richness and X-ray luminosity, highlighting in black clusters identified as contaminated. We exclude these objects from our analysis. In Section~\ref{subsec:selection_bias} we discuss the possible implications of this selection on our results.

\subsection{Galaxy number density profile}
\label{sec:number_density}

As showed in Section~\ref{sec:theory}, the Jeans analysis requires knowledge of the 3D number density profile $\nu (r)$ of the tracer population, i.e. the red sequence member galaxies. 
The absolute normalization of the galaxy number density profile has no impact on our analysis, because only the logarithmic derivative of $\nu (r)$ enters the Jeans equation (see equation~\ref{eq:jeans}). 
On the other hand, a radially dependent incompleteness in the velocity sample would lead to a modification of the shape of the  $\nu (r)$ profile, which would have an impact on our results. 
As the spectroscopic followup within SPIDERS will lead to a radially dependent incompleteness, we cannot simply adopt the spectroscopic sample to measure the number density profile of the tracer  population. We therefore rely on a study of the galaxy populations in 74 SZE selected  clusters from the SPT-SZ survey, imaged as part of the Dark Energy Survey Science Verification phase \citep{2017Hennig}. This study shows that the number density profile of the red sequence population is well fit by a Navarro, Frenk and White (NFW) model \citep[][]{1996NFW} out to radii of 4$r_\mathrm{200c}$, with a  concentration for cluster galaxies of $c_{\text{gal}} = 5.37^{+0.27}_{-0.24}$. No statistically significant redshift or mass trends were identified in the radial distribution of red sequence galaxies for $z>0.25$ and $M_\mathrm{200c}>4\times10^{14}M_\odot$. 
Therefore, we adopt the number density profile described by an NFW profile with the above-mentioned value  of $c_{\text{gal}}$ and a scale radius $r_{\text{s, gal}} = R_\mathrm{200c}/c_{\text{gal}}$. Implicit in this approach is the assumption that the dynamical properties of our spectroscopic sample are consistent with those of the red sequence galaxy population analyzed by \citet{2017Hennig}.

We examine the impact of this assumption on our results by performing our analysis on a range of concentrations, spanning the range $3.5 < c_{\text{gal}} < 6$. In Appendix~\ref{sec:appendix} we show that our results are not very sensitive to the choice of the concentration parameter. We will nevertheless further examine the impact of mismatch between the model and actual radial distribution of the tracer population in an upcoming study, in which we seek to improve the understanding of biases and scatter in dynamical mass estimators using mock observations of structure formation simulations (Capasso et al., in prep.).

%%%%%%%%%%%%%%%%%%%%%%%%%%%%%%%%%%%%%%
%%%% %%%                            Results                                 %%%%%%%%
%%%%%%%%%%%%%%%%%%%%%%%%%%%%%%%%%%%%%%

\begin{table} \centering
\caption{Priors assumed for our analysis. $\mathcal{U}(i, j)$ refers to a uniform flat prior in the interval $(i, j)$, while $\mathcal{N}(\mu, \sigma^{2})$ indicates a Gaussian distribution with mean $\mu$ and variance $\sigma^{2}$.} 
\begin{tabular}{ccccc}
\hline\\[-7pt]
$\rm{A}_{L_{\rm{X}}}$ & $\rm{B}_{L_{\rm{X}}}$ & $\gamma_{L_{\rm{X}}}$  & $\theta_{\beta}$ & $\sigma_{\ln L_{\rm{X}}^{\text{int}}}$ \\ [2pt]
\hline\\[-7pt]
$\mathcal{U}(0.1, 2)$ & $\mathcal{U}(1, 5)$ & $\mathcal{U}(-7, 2)$  &  $\mathcal{U}(0.01, 10)$ & $\mathcal{N}(0.27, 0.1^2)$ \\ [3pt]
\hline
\end{tabular}
\label{tab:Lxpriors}
\end{table}

\begin{table*} \centering
\caption{X-ray luminosity-mass-redshift scaling relation parameters and intrinsic scatter from this analysis and the literature. Parameters are as defined in equation~(\ref{eq:Lx-mass}), and include the Eddington and Malmquist biases. Results from this analysis are showed with statistical uncertainties, together with systematic mass uncertainties.  In the comparison to previous results, the amplitude $\mathrm{A}_{\rm{X}}$ column contains the luminosity at $M_\mathrm{200c}=3\times10^{14} M_\odot$ and  $z=0.16$.  Conversions have been made to $M_\mathrm{200c}$ and from $E(z)$ to $(1+z)$ where needed. Note also that each of these studies was performed on a different range of mass and redshift.} 
\begin{tabular}{lccccc}
\hline\\[-7pt]
Dynamical analysis using SPIDERS data & $\mathrm{A}_{\rm{X}}$ & $\mathrm{B}_{\rm{X}}$ & $\gamma_{\rm{X}}$ & $\mathrm{C}_{\rm{X}}$ & $\sigma_{\ln L_{\rm{X}}}^{\text{int}}$ \\ [2pt]
\hline\\[-7pt]
Baseline analysis & $0.62^{+0.05}_{-0.06}\pm0.06$ & $2.35^{+0.21}_{-0.18} \pm0.09 $ &  $-2.77^{+1.06}_{-1.05} \pm0.79$ & $2$ & $0.25^{+0.09}_{-0.09}$ \\ [3pt]
Combined analysis & $0.60^{+0.05}_{-0.06}$ & $2.01^{+0.09}_{-0.09}$  &  $-0.56^{+0.36}_{-0.36}$ & $2$ & $0.23^{+0.09}_{-0.08}$ \\ [3pt]
\hline\\[-7pt]
Previously published results & &  &  &  & \\[4pt]
\hline\\[-7pt]
SPT + \XMM\ \citep{2019Bulbul} & $0.58\pm0.09$ & $1.92\pm0.18$ & $0.004\pm0.50$  & $2$ & $0.27\pm0.10$ \\[3pt]
WL + RASS \citep{2018Nagarajan} & $0.42\pm0.27$ & $1.62\pm0.30$ & -- & $2$ & $0.75^{+0.19}_{-0.16}$ \\[3pt]
Chandra \citep{2017Giles} & $0.27\pm0.13$ & $1.96\pm0.24$ & -- & $2$ & $0.68\pm{0.11}$ \\[3pt]
Chandra + ROSAT \citep{2016Mantz} & $0.80\pm0.35$ & $1.35\pm0.06$ & $-0.65\pm0.38$  & $2.31\pm{0.06}$ & $0.42\pm{0.03}$ \\[3pt]
Chandra + ROSAT \citep{2009aVikhlinin} & $0.68\pm0.21$ & $1.63\pm0.15$ & -- & $1.85\pm{0.42}$ & $0.40\pm{0.04}$ \\[3pt]
{\gwpfont REXCESS} \citep{2009Pratt} & $1.04\pm0.09$ & $1.63\pm0.11$ & -- & $2.33$ & $0.41\pm{0.07}$ \\[3pt]
\hline\\[-7pt]
%\hline
\end{tabular}
\label{tab:results}
\end{table*}

\begin{figure}
\centering { 
   \includegraphics[scale=0.55]{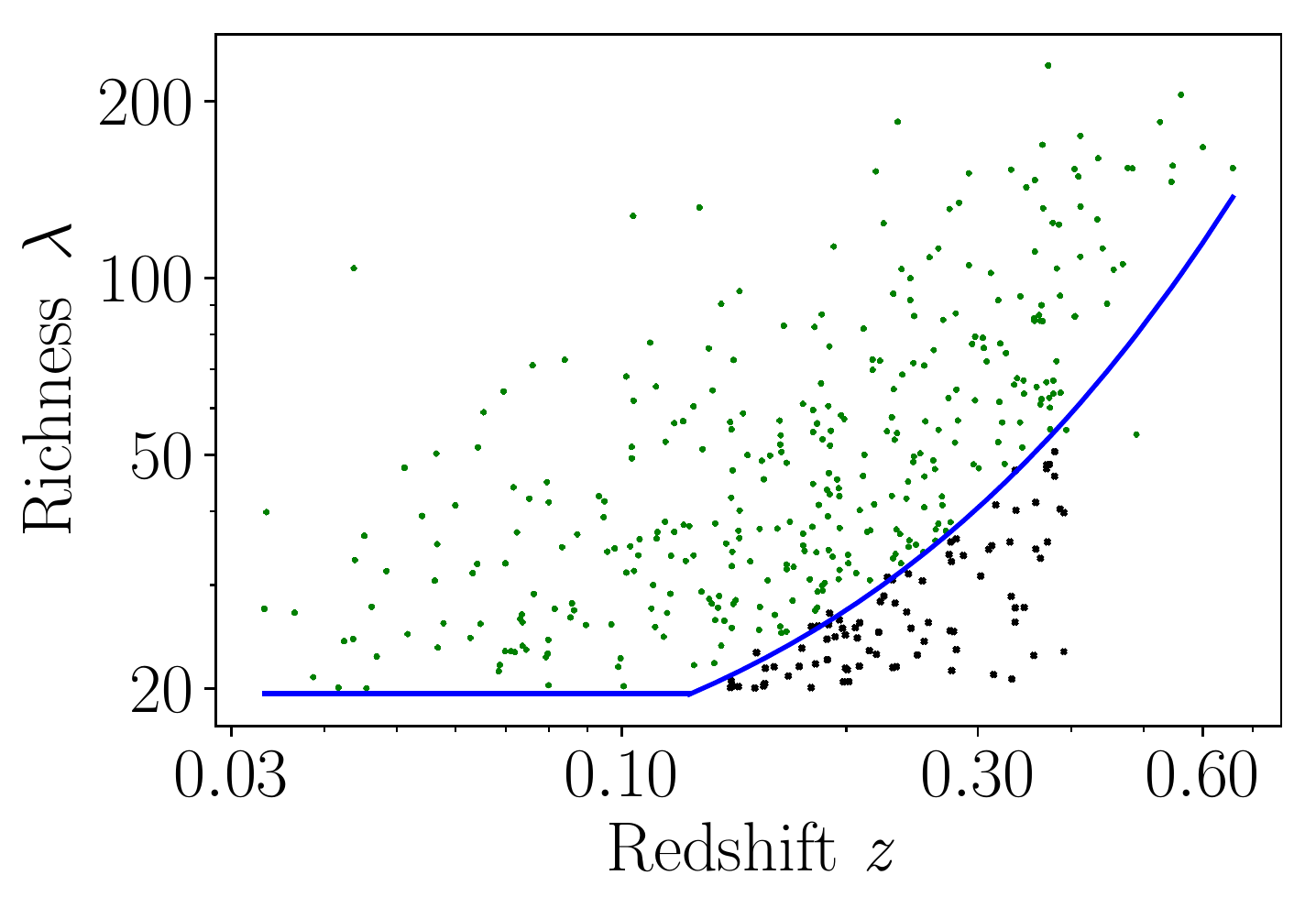} 
      }
\caption{Distribution of our cluster sample in richness and redshift. Black crosses represent the contaminated sources that we exclude from the sample. The richness sample selection is shown by the blue line.}
\label{fig:contamination_bias}
\end{figure}

\begin{figure}
\centering { 
   \includegraphics[scale=0.42]{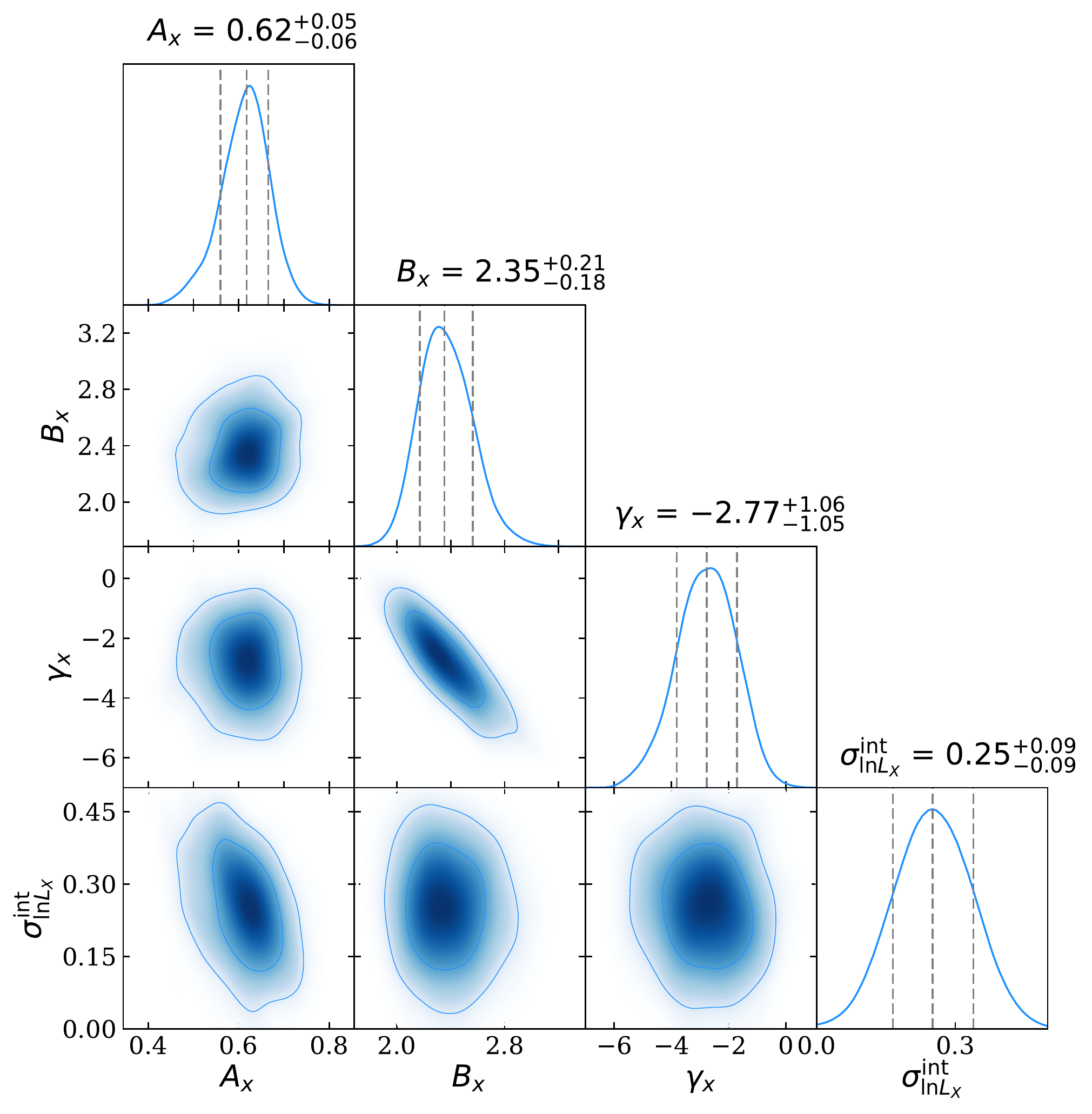} 
      }
\caption{Posterior distribution of the scaling relation parameters. Contours show the 1$\sigma$, 2$\sigma$, and 3$\sigma$ confidence regions.}
\label{fig:corner_plot}
\end{figure}

\section{Results}
\label{sec:results}

This section is dedicated to the results of the dynamical analysis. In the first subsection we present the method used to calibrate the $L_{\text{X}}$-mass relation, and the results we obtain. We end the section with a comparison of our findings to those from previous studies, and we discuss the impact of the choice of the priors on our results.

\subsection{Fitting Procedure}
\label{subsec:fit}

We model the relation between the X-ray luminosity, mass and redshift as
\be
\label{eq:Lx-mass}
\dfrac{L_{\rm{X}}}{(10^{44} \rm{erg\,s^{-1}})} \! = \!
\text{A}_{\rm{X}} \!
\left( \! \dfrac{M_{\text{200c}} \!} {\! M_{ \text{piv}} \!} \right)^{\!\text{B}_{\rm{X}}} \!\!
 \left(\!  \dfrac{E(z) \!}{\! E(z_{\rm{piv}})\!}   \right)^{\!\text{C}_{\rm{X}}}  \!\!
 \left(\! \dfrac{1\!+\!z \!}{ \! 1\!+\!z_{ \rm{piv}} \!} \right)^{\!\gamma_{\rm{X}}} \!\!, \!\!
\ee
\\
where $\text{A}_{\rm{X}}$, $\text{B}_{\rm{X}}$, and $\gamma_{\rm{X}}$ are the amplitude, the mass slope and the redshift evolution slope. In this formulation, the redshift trend is expressed as both a function of $z$, and of the Hubble parameter $H(z) = H_{0} E(z)$. In this analysis, we fix $\text{C}_{\rm{X}} = 2$. In a flat $\Lambda$CDM Universe, $E^2(z) = \Omega_{\rm{m}}(1+z)^3 + \Omega_{\Lambda}$ at late times. Therefore, we explicitly parametrize the cosmological dependence of the redshift evolution, while modeling departures  from the self-similar evolution with a function $(1+z)^{\gamma_{\rm{X}}}$. Similar forms have been previously adopted to study the redshift and mass trends of the $L_{\rm{X}}$-mass relation  \citep[e.g.,][]{2019Bulbul}.
The redshift and mass pivot points are set to be $z_{ \text{piv}} = 0.16$ and $M_{ \text{piv}}=3 \times 10^{14}M_\odot$, respectively, corresponding to the median mass and redshift of our sample, where the $M_{ \text{piv}}$ has been chosen a posteriori, after a first run of the analysis.
We adopt a log-normal intrinsic scatter in $L_{\rm{X}}$ at fixed mass, $\sigma_{\ln L_{\rm{X}}}^{\text{int}}$. 

We follow the fitting framework presented in \citetalias{2019bCapasso}. Given the set of parameters $\textbf{\textit{p}}$, containing the 4 scaling relation parameters ($A_{\rm{X}}, B_{\rm{X}}, \gamma_{\rm{X}}$, $\sigma_{\ln\lambda}^{\text{int}}$) and the anisotropy model parameter $r_{\beta}$, we calculate an initial mass $M_{\text{200c,obs}}$ using the scaling relation presented in Eq.~\ref{eq:Lx-mass}. We then use the method of \citet{2011Mortonson} to estimate the Eddington bias correction caused by the interplay of the cluster mass function and scatter of the scaling relation.  We assume the variance on the log-normal mass-observable relation to be $\sigma^2_{\ln{M}} = (1 / B_{\rm{X}} \cdot \sigma_{\ln L_{\rm{X}}})^{2} $, where 
\be
\label{eq:scatter}
\sigma_{\ln L_{\rm{X}}}^{2} = \left( \dfrac{\Delta L_{\rm{X}}}{L_{\rm{X}}} \right)^2 +  {\sigma_{\ln L_{\rm{X}}}^{\text{int}}}^{2}    ,
\ee
with $\Delta L_{\rm{X}}$ being the $L_\mathrm{X}$ measurement uncertainty divided by the observed luminosity. Assuming that the variance $\sigma^2_{\ln{M}}$ is small compared with the scale over which the local slope $\Gamma$ of the mass function changes, the posterior mass distribution is a log-normal of the same variance $\sigma^2_{\ln{M}}$ with the mean shifting as $\ln <M_{\text{200c,true}}> = \ln <M_{\text{200c,obs}}> + \Gamma\sigma_{\ln M}^{2} $. 
We adopt this mass as input in MAMPOSSt, evaluating for each cluster the likelihood distribution in projected phase space. We combine the likelihoods calculated for each member galaxy in that cluster, such that the $i-$th term in the likelihood $\mathcal{L}_{i}$ contains the probability of observing the $i-$th cluster at redshift $z_{i}$, with mass $M_{\text{200c,true, i}}$ and X-ray luminosity $L_{\rm{X}, i}$, and the phase space of its member galaxies (clustercentric radii $R^{j}$ and rest-frame velocities $v^{j}_{rf}$ of each $j-$th galaxy), given the scaling relation parameters $\text{A}_{\rm{X}}$, $\text{B}_{\rm{X}}$ and $\gamma_{\rm{X}}$, the anisotropy parameter $r_{\beta}$, and the intrinsic scatter $\sigma_{\ln L_{\rm{X}}}^{\text{int}}$:
\be
\label{eq:lik_Lx}
\mathcal{L}_{i}=\prod_{j \in gal} \, \mathcal{L} (R^{j}, v^{j}_{rf}, L_{\rm{X}, i}, z_{i} \mid  \textbf{\textit{p}}). 
\ee
The maximum likelihood solutions are obtained using the \textsc{newuoa} software \citep{newuoa}. Priors on the parameters are assumed as follows (see Table~\ref{tab:Lxpriors}): flat for the scaling relation parameters and for the anisotropy parameter, gaussian for the intrinsic scatter (mean $\mu=0.27$ and variance $\sigma^{2}={0.1}^2$, from \citet{2019Bulbul}).

The final likelihood for the total sample, for each set of scaling relation parameters~$\textbf{\textit{p}}$, will then be obtained by combining the likelihoods for all the single clusters:
\be
\mathcal{L}  = \prod_{i \in clus} \mathcal{L}_{i} .
\ee

This procedure is carried on separately for each anisotropy profile model (see Section~\ref{sec:profiles}). The posterior parameter distributions obtained from the different anisotropy models are then combined by means of the Bayesian model averaging technique, effectively marginalizing over the uncertainties in the orbital anisotropy (see discussion in Section~\ref{sec:Bayes}).

\subsection{Systematic Effects}
\label{sec:systematics}

This section is dedicated to estimating the systematic errors entering our analysis and their impact on the best fit parameter uncertainties.

\subsubsection{Selection bias}
\label{subsec:selection_bias}

As described in Section~\ref{sec:CODEX} (see also the bottom panel of Figure~\ref{fig:histo}), the sample analyzed in this work is mainly flux-limited, with a further richness selection. This selection could introduce a bias into our analysis. To estimate the impact of this systematic bias on our results, we estimate its effects on a mock sample.

Starting from the halo mass function \citep{2008Tinker}, we create a large mock catalogue, computing the number of expected clusters as a function of halo mass and redshift ($\sim3.6\times10^6$ clusters). 
We draw a Poisson realization of this dataset, obtaining a mass selected sample made of $\sim800$ clusters (doubling the observed dataset), with $M_{\text{200c}} \ge 5\times 10^{13} $ and $ 0.05 \le z \le 0.66$. 
To each cluster we assign a luminosity sampled from a Gaussian distribution centered on the X-ray luminosity predicted by our analysis using one anisotropy model, namely the constant anisotropy model, and scatter given by $\sigma_{\ln L_{\rm{X}}}^{\text{int}}$. 
The input values for the scaling relation, which thus slightly differ from the values listed in Table~\ref{tab:results}, are as follows: $A_{\rm{X}} = 0.54$, $\text{B}_{\rm{X}} = 2.37$ and $\gamma_{\rm{X}} = -2.41$. 
To create the sample of member galaxies for each cluster, we run MAMPOSSt on a grid of velocities and radii, fixing the galaxy number density profile to that described in Section~\ref{sec:number_density}. From the likelihood we derive the probability density of observing an object at a certain projected phase space location \citep[see equation~11,][]{2013MAMPOSSt}, drawing a random number of galaxies from the observed distribution of member galaxies.
We fit this mock sample following the procedure described in Section~\ref{subsec:fit}, recovering best fit parameters consistent with the input values.
As a second step, we convert the X-ray luminosity in flux using the equation $L_{\rm{X}}/(4\pi\rm{D_{L}}^2)$, where $\rm{D_{L}}$ is the luminosity distance at fixed cosmology and redshift. Finally, we calculate richness using the scaling relation calibrated in  \citetalias{2019bCapasso}. 
To estimate the effect of the selection bias, we impose the same cuts
that are applied to the observed sample, i.e. a flux cut at $F_{\rm{X}} > 1.4 \times 10^{-13} [erg/s/cm^2]$, and a richness cut at $\lambda > 20$ combined with a redshift dependent richness cut due to the removal of random superposition (see blue line in Figure~\ref{fig:contamination_bias}). 
Performing our analysis on this mock sample, consisting of $\sim800$ clusters and $\sim22,400$ member galaxies, we recover the following constraints on the scaling relation parameters: $A_{\rm{X}} = 0.53^{+0.02}_{-0.02}$, $\text{B}_{\rm{X}} = 2.45^{+0.09}_{-0.09}$ and $\gamma_{\rm{X}} = -2.58^{+0.52}_{-0.58}$. These values are less than $1\sigma$ away from the input parameters, and small compared to the statistical uncertainties reported in Table~\ref{tab:results}. We conclude our results are not significantly affected by a selection bias.

\subsubsection{Systematics in MAMPOSSt mass estimates}

Another systematic effect we need to take into account is the one associated with the dynamical mass measurements themselves. To estimate this additional systematic uncertainty we employ the findings of \citet[][]{2013MAMPOSSt}, recovered by analyzing runs of the MAMPOSSt code on numerical simulations. Using particles lying within a sphere of $r_{100}$ around the halo center, they show that the estimated value of the cluster virial radius $r_\mathrm{200c}$ is biased at $ \leq 3.3\%$ \citep[see Table 2,][]{2013MAMPOSSt}. We thus adopt a Gaussian systematic uncertainty on the virial mass $M_\mathrm{200c}$ of $\sigma = 10\%$. As the \citet{2013MAMPOSSt} analysis does not explore mass or redshift trends in these biases, we apply the entire uncertainty to the normalization parameter $\mathrm{A}_{\rm{X}}$. In a future analysis, we plan to explore the mass and redshift dependence of the systematic uncertainties in dynamical mass estimates from a Jeans analysis on numerical simulations.

%%%%%%%%%%%%%%%%%%%%%%%%%%%%%%%%%%%%%% 

\begin{table} 
\centering
\caption{Impact of the number of member galaxies on the luminosity-mass-redshift scaling relation parameters, defined in equation~(\ref{eq:Lx-mass}). The uncertainties on the results are statistical, corresponding to 68 per cent confidence intervals.} 
\begin{tabular}{cccc}
\hline\\[-7pt]
Number of cluster  & $\mathrm{A}_{\rm{X}}$ & $\mathrm{B}_{\rm{X}}$ & $\gamma_{\rm{X}}$  \\ [2pt]
member galaxies &  & &   \\ [2pt]
\hline\\[-7pt]
$N_\mathrm{mem}\ge1$ & $0.64^{+0.05}_{-0.06}$ & $2.20^{+0.18}_{-0.16}$  &  $-1.19^{+0.73}_{-0.80}$ \\ [3pt]
$N_\mathrm{mem}\ge3$ & $0.64^{+0.05}_{-0.06}$ & $2.24^{+0.19}_{-0.17}$  &  $-1.39^{+0.83}_{-0.89}$ \\ [3pt]
$N_\mathrm{mem}\ge5$ & $0.64^{+0.05}_{-0.06}$ & $2.27^{+0.17}_{-0.16}$  &  $-1.75^{+0.81}_{-0.83}$ \\ [3pt]
$N_\mathrm{mem}\ge10$ & $0.62^{+0.05}_{-0.06}$  & $2.35^{+0.21}_{-0.18}$  &  $-2.77^{+1.06}_{-1.05}$  \\ [3pt]
$N_\mathrm{mem}\ge15$ & $0.59^{+0.05}_{-0.06}$ & $2.38^{+0.25}_{-0.21}$  &  $-2.52^{+1.14}_{-1.30}$ \\  [3pt]
\hline
\end{tabular}
\label{tab:Nmem}
\end{table}

\subsubsection{Impact of the number of member galaxies}
\label{sec:nmem}

As mentioned in Section~\ref{sec:interloperrejection}, we apply a cut on the number of spectroscopic members per cluster, $N_{\text{mem}}\ge10$. This choice derives from the concern that below a certain number of cluster members even the mean redshift of the cluster becomes uncertain and the dynamical information becomes too noisy to be reliable for a scaling relation reconstruction.  Following \citetalias{2019bCapasso}, we estimate the impact of this cut on our results.  

Table~\ref{tab:Nmem} lists the constraints on the best fit parameters for varying values of $N_{\text{mem}}$, from 1 to 15.  We note that, as the BCG has already been excluded, clusters with  $N_{\text{mem}}=1$ actually have two measured spectroscopic redshifts. As in \citetalias{2019bCapasso}, this cut does not significantly affect the normalization $\mathrm{A}_{\rm{X}}$ and the mass trend parameter $\mathrm{B}_{\rm{X}}$. On the other hand, the redshift trend parameter $\mathrm{\gamma}_{\rm{X}}$ changes considerably.  The value of $\mathrm{\gamma}_{\rm{X}}$ starts converging when including only clusters with at least 10 spectroscopic members, justifying the cut imposed on our sample.  

The strong dependence of $\gamma_{\rm{X}}$ on the number of galaxies could indicate an additional source of systematic uncertainty. In the discussion presented in \citetalias{2019bCapasso}, we highlight that the distribution of clusters with $N_{\text{mem}}<10$ extends to higher redshifts, representing a qualitatively different population of objects. To assess whether the trend in $\gamma_{\rm{X}}$ represents the true redshift trend or is a sign of a systematic in the limit of low spectroscopic sampling, further exploration with a larger high $z$ spectroscopic sample is needed. In the meantime, we use this apparent trend to estimate a systematic uncertainty on the scaling relation parameters. As in \citetalias{2019bCapasso}, we define this uncertainty as half the full range of variation in the value of the parameter, $\sigma_{\text{sys},\gamma_{\rm{X}}}={\Delta |\gamma_{\rm{X}} |\over 2}=0.79$. 
We also estimate this factor for the mass trend parameter, $\sigma_{\text{sys},\mathrm{B}_{\rm{X}}}={\Delta |\mathrm{B}_{\rm{X}} |\over 2}=0.09$. For the amplitude parameter, on the other hand,  the shift is small compared to the 10\% systematic uncertainty described at the beginning of this section. These systematic uncertainties are included in results listed in  Table~\ref{tab:results}.

\subsection{Parameter constraints}
\label{subsec:constraints}

The resulting posteriors of our scaling relation parameters are summarized in Table~\ref{tab:results}. The uncertainties are statistical, together with the additional 10\% systematic uncertainty described above. 
Figure~\ref{fig:corner_plot} shows the corresponding joint parameter constraints. We find that galaxy clusters with mass $M_{\text{200c}} = 3 \times 10^{14} M_{\odot}$  at $z = 0.16$ have a mean X-ray luminosity $L_{\rm{X}} = 0.62^{+0.05}_{-0.06} \times 10^{44}$, and scale with mass and redshift as $\text{B}_{\rm{X}} = 2.35^{+0.21}_{-0.18}$ and $\gamma_{\rm{X}} = -2.77^{+1.06}_{-1.05}$ respectively. The posterior distribution of the intrinsic scatter is consistent with that of the prior.
In the following section we compare our calibration of the $L_{\rm{X}}$-mass relation to previous results from the literature.

\begin{figure}
\centering { 
   \includegraphics[scale=0.52]{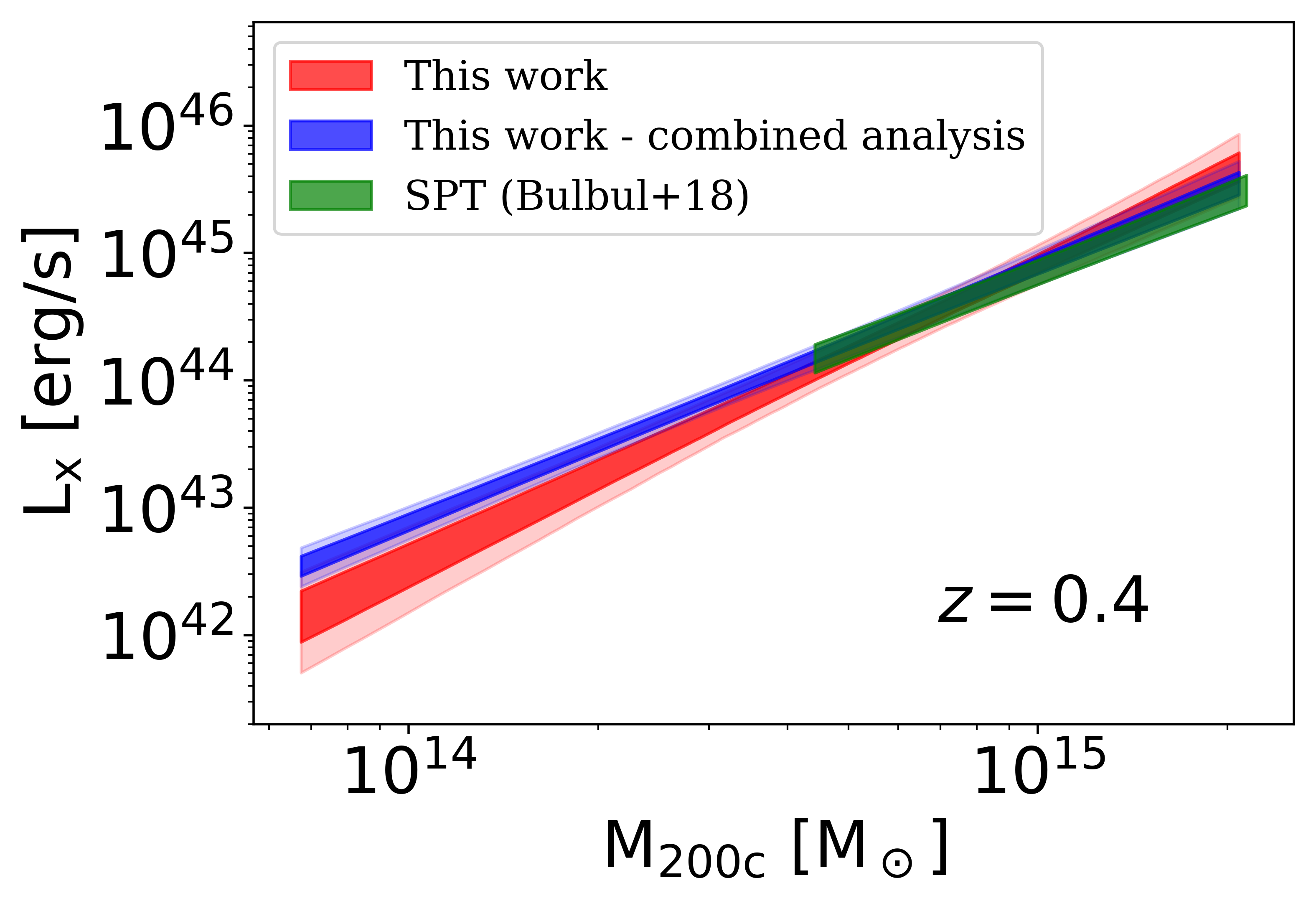} 
      }
\caption{Best fit model for our X-ray luminosity-mass relation (in red), evaluated at the redshift $z = 0.4$, 
compared to the \citet{2019Bulbul} measurements. In blue we show the results from a combined analysis of the two results. Shaded regions correspond to the 1$\sigma$ and 2$\sigma$ confidence regions. For the  \citet{2019Bulbul} results we only show the 1$\sigma$ confidence area.}
\label{fig:Lx_comp}
\end{figure}

\begin{figure}
\centering 
\includegraphics[scale=0.52]{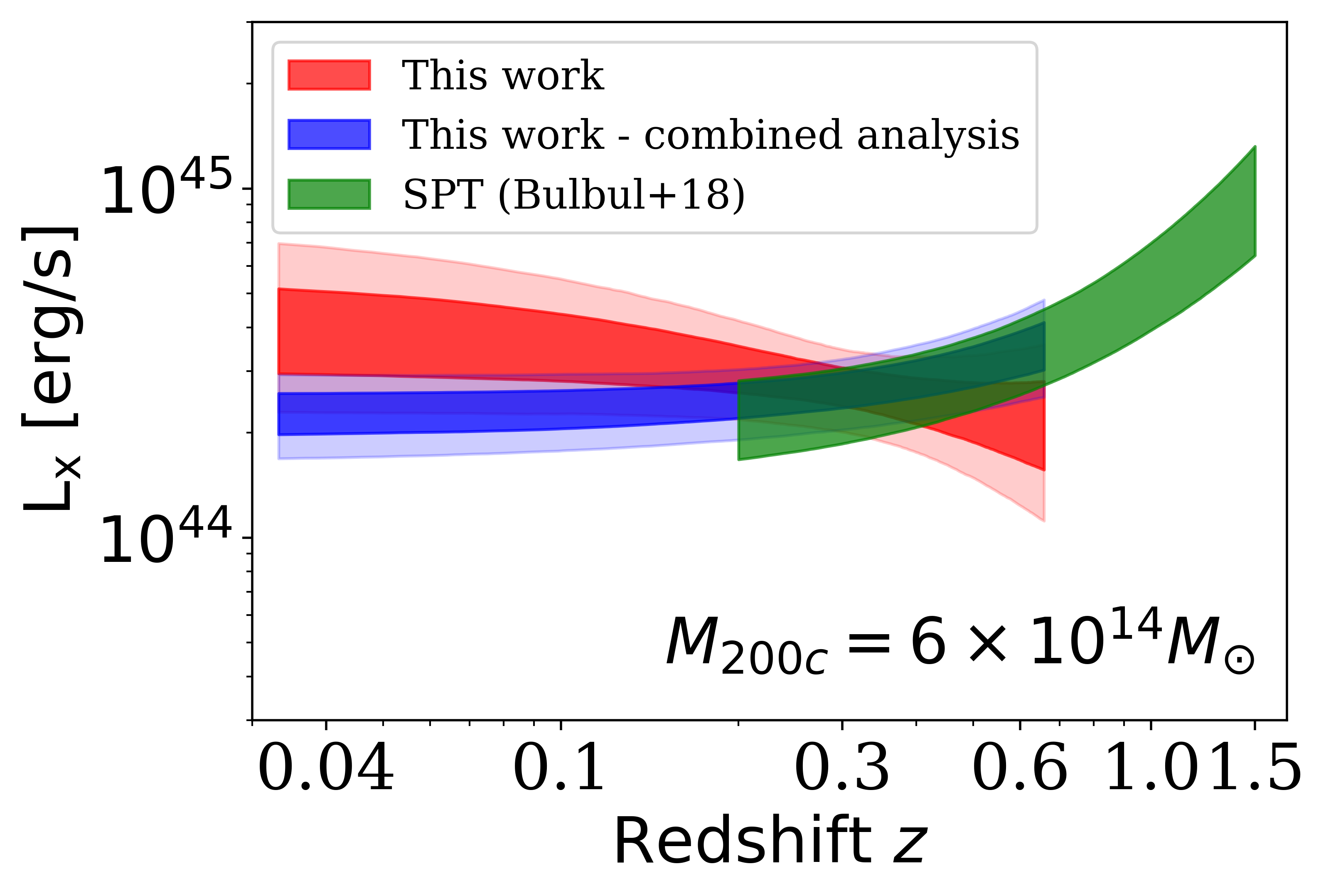}
\caption {Best fit model for our X-ray luminosity-mass relation (in red), evaluated at the mass 
$M_{\rm{200c}} = 6 \times 10^{14} M_{\odot}$, compared to the \citet{2019Bulbul} measurements. In blue we show the results from a combined analysis of the two results. Shaded regions correspond to the 1$\sigma$ and 2$\sigma$ confidence regions. For the  \citet{2019Bulbul} results we only show the 1$\sigma$ confidence area.}
\label{fig:Lxz_evol}
\end{figure}

\begin{figure}
\centering { 
   \includegraphics[scale=0.42]{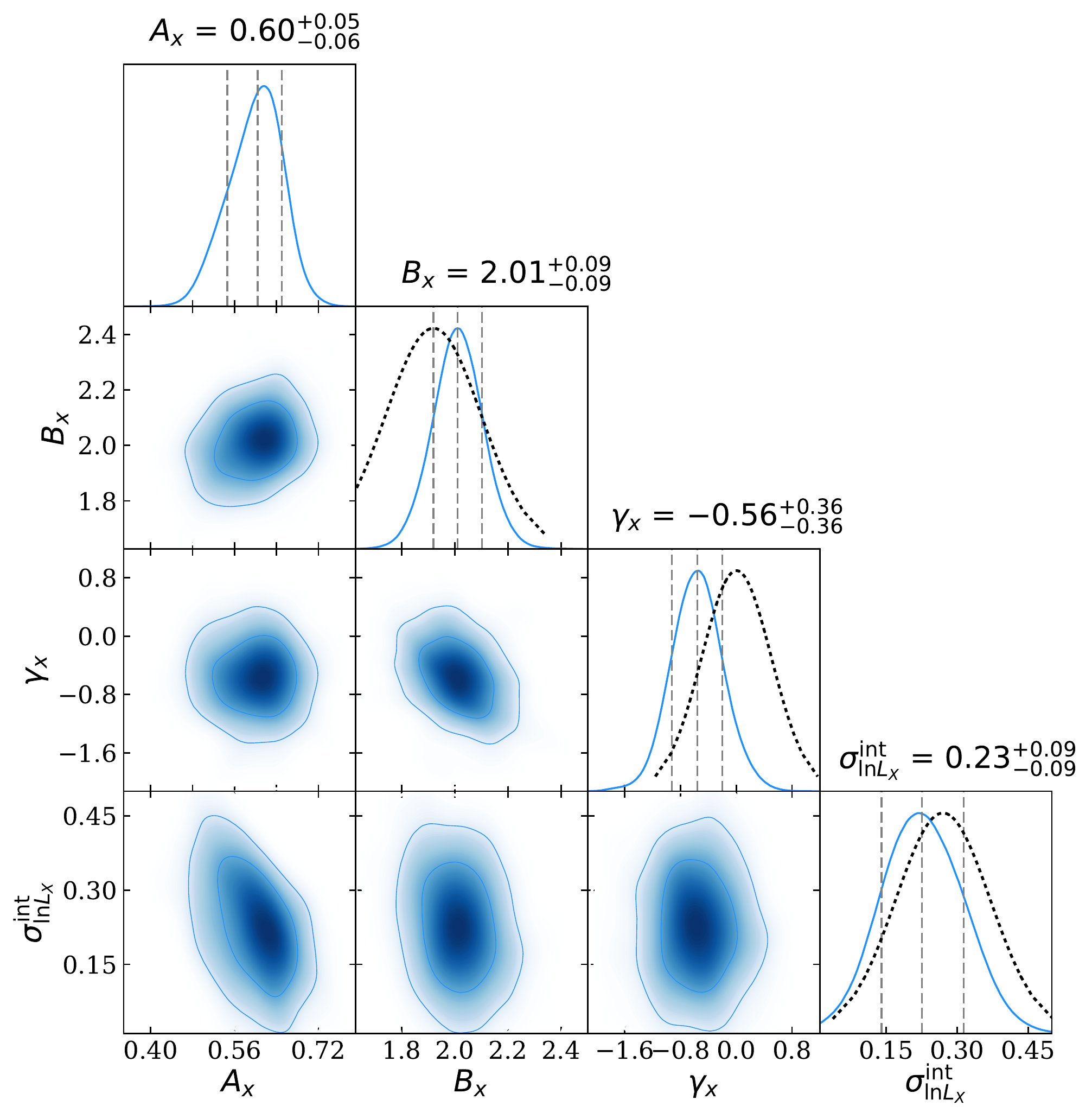} 
      }
\caption{Posterior distribution of the scaling relation parameters for the combined analysis. Contours show the 1$\sigma$, 2$\sigma$, and 3$\sigma$ confidence regions, also highlighted by the vertical dashed lines. The prior distributions for the mass and redshift trends and the intrinsic scatter are taken from \citet{2019Bulbul} and shown in black. The prior on the normalization is the flat one showed in Table~\ref{tab:Lxpriors}.}
\label{fig:corner_plot_combined}
\end{figure}

\subsection{Comparison to previous results}
\label{subsec:comp}

Table~\ref{tab:results} summarizes the parameter estimates and uncertainties for our analysis and the \citet{2019Bulbul} comparison results. To make this comparison, we scale the measurements from \citet{2019Bulbul} to the redshift $z_{\text{piv}} = 0.16$, and mass $M_{\text{piv}} = 3 \times 10^{14} M_{\odot}$, using the published best fit redshift and mass trends. The mass conversion from $M_{\rm{500c}}$ to $M_{\rm{200c}}$ is carried out using \textsc{Colossus}, an open-source python package for calculations related to cosmology \citep{2017Diemer}. Moreover, the analysis performed in \citet{2019Bulbul} is also based on luminosities extracted within $r_{500}$ (referred to as core-included), but in a rest-frame band of [0.5-2]~keV. We thus estimate a factor 1.6 to be applied to their amplitude.

Figures~\ref{fig:Lx_comp} and~\ref{fig:Lxz_evol} show the mass and redshift trends of the X-ray luminosity,  where for the redshift trend we correct the data points to the mass $M_\text{200c}=6\times10^{14}M_\odot$ and for the mass trend we move the data points to the redshift $z=0.4$. These values have been chosen as clusters with such mass at this redshift are present in both our dataset and the one analyzed by \citet{2019Bulbul}. This means that these are not the places where our constraints are tightest, but the ones where the comparison between the works is justifiable. 
The best fit model for the $L_{\rm{X}}-M_{\text{200c}}$ relation is shown in red, with shaded $1\sigma$ and $2\sigma$ confidence regions. For the results from \citet[][in green]{2019Bulbul}, we show only the $1\sigma$ confidence region. We limit the redshift range to that analyzed in each work.  

Our mass trend shows good agreement with the results obtained by \citet{2019Bulbul}, based on \XMM\ X-ray observations of an SZE selected sample from the South Pole Telescope 2500~deg$^2$ SPT-SZ survey (Fig.~\ref{fig:Lx_comp}). We also find a good agreement with the results reported from weak-lensing derived masses of an X-ray selected sample \citep[APEX-SZ;][]{2018Nagarajan}. Additionally, our mass trend is consistent with that found by \citet{2017Giles}, obtained through an analysis of galaxy clusters observed with \textit{Chandra}, and \citet{2009aVikhlinin}, based on \textit{Chandra} observations of samples derived from the ROSAT All-Sky survey. However, we find a steeper $\text{B}_{\rm{X}}$ compared to that reported in \citet{2016Mantz}, also based on \textit{Chandra} and ROSAT data. We also compare our mass trend with that found by \citet{2009Pratt}, based on the Representative \XMM\ Cluster Structure Survey ({\gwpfont REXCESS}) dataset. We find our scaling relation to be somewhat steeper (at a $3.3\sigma$ level) than their [0.1-2.4]~keV band scaling relation, corrected for the Malmquist bias. Overall, our study recovers a steeper than self-similar mass trend, in agreement with most previously published analyses.

Our constraint on the redshift trend of the $L_{\rm{X}}-M_{\text{200c}}-z$ relation, on the other hand, suggests a stronger negative evolution than found by \citet{2019Bulbul} (Fig.~\ref{fig:Lxz_evol}). However, we note that the redshift range probed by \citet{2019Bulbul} is higher and complementary to that covered by our sample. In Section~\ref{subsec:combined} we describe the results obtained by combining the two samples.
Our redshift trend (both from our baseline analysis and the combined one) is also in good agreement with the value of $\gamma_{\rm{X}}$ found by \citet{2016Mantz}. We also note that all the results from the literature we cited assume a self-similar evolution of the form $E(z)^{2}$, apart from \citet{2016Mantz},  \citet{2009aVikhlinin} and \citet{2009Pratt}. For a discussion of the expected self-similar trends in mass and redshift, we refer the reader to \citet{2019Bulbul}.

\subsection{Combined analysis}
\label{subsec:combined}

Our sample and the one analyzed by \citet{2019Bulbul} cover complementary ranges of mass and redshift.  In particular, the SPT selected cluster sample extends to higher redshift, and is therefore helpful in constraining the redshift evolution parameter of the scaling relation. Therefore, we perform a ``combined" analysis by adopting the priors on the mass and redshift trends found by  \citet{2019Bulbul}. Figure~\ref{fig:corner_plot_combined} shows the posterior distribution of the scaling relation parameters, together with the prior distributions for the mass and redshift trends and the intrinsic scatter. We note that the two sets of distributions are in agreement, allowing us to perform this joint analysis. The prior on the normalization is the same flat prior used for the CODEX-only analysis (see Table~\ref{tab:Lxpriors}). The results are listed in Table~\ref{tab:results}. Figures~\ref{fig:Lx_comp}, \ref{fig:Lxz_evol} and \ref{fig:corner_plot_combined} all demonstrate that the results of the combined analysis are fully consistent with the \citet{2019Bulbul} ones, showing a shallower mass trend and a higher value of the redshift trend.

\section{Conclusions}
\label{sec:conclusions}

We present the calibration of the X-ray luminosity--mass--redshift relation using galaxy dynamical information from a sample of 344 CODEX galaxy clusters. These systems are X-ray selected clusters from RASS that have red-sequence selected redMaPPer optical counterparts within a search radius of 3$'$.  The sample is cleaned of random superpositions using an $f_\mathrm{cont}=0.05$ cut \citep{2019Klein}, which reduces the contamination from an initial $\sim$25\% to a target 5\%. The cluster sample we analyze has redshifts up to $z\sim0.66$, optical richness $\lambda \ge 20$, and spans an X-ray luminosity range $ 4.5\times10^{42} \leq  L_{\rm{X}}/(\rm{erg\,s}^{-1})  \leq 3.2\times10^{45} $. The spectroscopic follow-up has been obtained from the SPectroscopic IDentification of eRosita Sources (SPIDERS) survey, resulting in a final sample of 6,658 red member galaxies.

We perform a Jeans analysis based on the code MAMPOSSt \citep{2013MAMPOSSt}. For each individual cluster, we extract the likelihood of consistency between the projected phase space distribution of the cluster members with measured redshifts and the modeled projected distribution for a cluster at redshift $z$, luminosity $L_{\rm{X}}$, and inferred mass $M_\text{200c}$. 
We adopt an NFW profile for the red galaxy tracer population with concentration $c=5.37$ \citep[][and Section~\ref{sec:number_density}]{2017Hennig}, and employ five different velocity dispersion anisotropy profiles.  We combine luminosity-mass relation posterior parameter distributions from the different anisotropy models by performing Bayesian model averaging, allowing us to marginalize over the orbital anisotropy of the spectroscopic galaxy population.

The scaling relation is modeled as $L_{\rm{X}} \propto \text{A}_{\rm{X}}M_{\text{200c}}^{\text{B}_{\rm{X}}} E(z)^2 (1+z)^{\gamma_{\rm{X}}}$ (equation~\ref{eq:Lx-mass}). We correct for the Eddington bias by implementing the method described in \citet{2011Mortonson}, which provides an estimate of the mean mass shift due to the log-normal mass observable relation scatter (equation~\ref{eq:scatter}) together with the measurement uncertainties on the X-ray luminosity. We also correct for the Malmquist bias, after evaluating its effect on a mock sample.

Results are showed in Table~\ref{tab:results}. For clusters of mass $M_{\text{piv}} = 3 \times 10^{14} M_{\odot}$, at redshift $z_{\text{piv}} = 0.16$, we find the following constraints on the scaling relation parameters:
\be 
\begin{split}
\mathrm{A}_{\rm{X}} = & 0.62^{+0.05}_{-0.06}\pm0.06,\\
\mathrm{B}_{\rm{X}} = & 2.35^{+0.21}_{-0.18}\pm0.09,\\
\gamma_{\rm{X}} = & -2.77^{+1.06}_{-1.05}\pm0.79,
\end{split}
\ee
where we quote systematic uncertainties for all the parameters. The amplitude uncertainty of 10\% comes from an estimate of the dynamical mass systematic uncertainty, applied to the scaling relation amplitude $\mathrm{A}_{\rm{X}}$ \citep[see study of systematics in][]{2013MAMPOSSt}.

Our results on the mass trend of the scaling relation are steeper, but statistically consistent (within $3\sigma$) with some previous literature results \citep{2019Bulbul, 2018Nagarajan, 2017Giles, 2009aVikhlinin}. However, we find mild disagreement with results from \citet{2009Pratt} (at $3.3\sigma$ level) and large departures from the \citet{2016Mantz} mass trend. 

We examine the redshift trend of the $L_{\rm{X}}$--mass scaling relation, finding a stronger negative, non-self-similar evolution of $L_{\rm{X}}$ with redshift with respect to the \citet{2019Bulbul} results. We explore this result by performing our analysis fixing the priors on our mass and redshift slopes to those adopted in \citet{2019Bulbul}. We recover the following set of parameters:
 \be 
\begin{split}
\mathrm{A}_{\rm{X}} = & 0.60^{+0.05}_{-0.06},\\
\mathrm{B}_{\rm{X}} = & 2.01^{+0.09}_{-0.09},\\
\gamma_{\rm{X}} = & -0.56^{+0.36}_{-0.36}. 
\end{split}
\ee
We note that the redshift trend has shifted to significant higher values, being consistent with the self-similar evolution and with previous studies.

This work, together with \citetalias{2019bCapasso}, shows the potential of dynamical masses in deriving mass--observable relations even in the limit of a small number of cluster members. This very promising result will be extremely useful in the context of future spectroscopic surveys like DESI \citep{2013Levi}, 4MOST \citep{dejong12}, Euclid \citep{2011Laureijs}, and the SDSS IV ``Black Hole Mapper" program \citep{2017Kollmeier}, which will focus also on the optical characterization of eROSITA X-ray sources. Performing a dynamical analysis on numerical simulations will enable significant improvements in the assessment of further systematic uncertainties, such as the impact of residual interlopers in our sample, departures from virial equilibrium, and variation of the velocity anisotropy profile (Capasso et al., in prep.).

\section*{ACKNOWLEDGMENTS}
The Munich group acknowledges the support by the DFG Cluster of Excellence ``Origin and Structure of the Universe'', the MPG faculty fellowship program and the Ludwig-Maximilians-Universit\"at Munich. RC acknowledges participation in the IMPRS on Astrophysics at the Ludwig-Maximilians University and the associated financial support from the Max-Planck Society. RC and VS acknowledge support from the German Space Agency (DLR) through \textit{Verbundforschung} project ID 50OR1603. AS is supported by the ERC-StG ``ClustersXCosmo", grant agreement 716762, and by the FARE-MIUR grant ``ClustersXEuclid'' R165SBKTMA. RC acknowledge financial support from the ERC-StG ``ClustersXCosmo", grant agreement 716762.  AB acknowledges the hospitality of the LMU and partial financial support from PRIN-INAF 2014 ``Glittering kaleidoscopes in the sky: the multifaceted nature and role of Galaxy Clusters?", P.I.: Mario Nonino. 

Funding for the Sloan Digital Sky Survey IV has been provided by the Alfred P. Sloan Foundation, the U.S. Department of Energy Office of Science, and the Participating Institutions. SDSS-IV acknowledges support and resources from the Center for High-Performance Computing at the University of Utah. The SDSS web site is www.sdss.org.

SDSS-IV is managed by the Astrophysical Research Consortium for the Participating Institutions of the SDSS Collaboration including the Brazilian Participation Group, the Carnegie Institution for Science, 
Carnegie Mellon University, the Chilean Participation Group, the French Participation Group, Harvard-Smithsonian Center for Astrophysics, Instituto de Astrof\'isica de Canarias, The Johns Hopkins University, Kavli Institute for the Physics and Mathematics of the Universe (IPMU) / 
University of Tokyo, the Korean Participation Group, Lawrence Berkeley National Laboratory, 
Leibniz Institut f\"ur Astrophysik Potsdam (AIP),  
Max-Planck-Institut f\"ur Astronomie (MPIA Heidelberg), 
Max-Planck-Institut f\"ur Astrophysik (MPA Garching), 
Max-Planck-Institut f\"ur Extraterrestrische Physik (MPE), 
National Astronomical Observatories of China, New Mexico State University, 
New York University, University of Notre Dame, 
Observat\'ario Nacional / MCTI, The Ohio State University, 
Pennsylvania State University, Shanghai Astronomical Observatory, 
United Kingdom Participation Group,
Universidad Nacional Aut\'onoma de M\'exico, University of Arizona, 
University of Colorado Boulder, University of Oxford, University of Portsmouth, 
University of Utah, University of Virginia, University of Washington, University of Wisconsin, 
Vanderbilt University, and Yale University.

%%%%%%%%%%%%%%%%% APPENDICES %%%%%%%%%%%%%%%%%%%%%

\appendix

\section{Test of fixed $c_{\rm{gal}}$}
\label{sec:appendix}

\begin{figure}
\centering { 
   \includegraphics[scale=0.42]{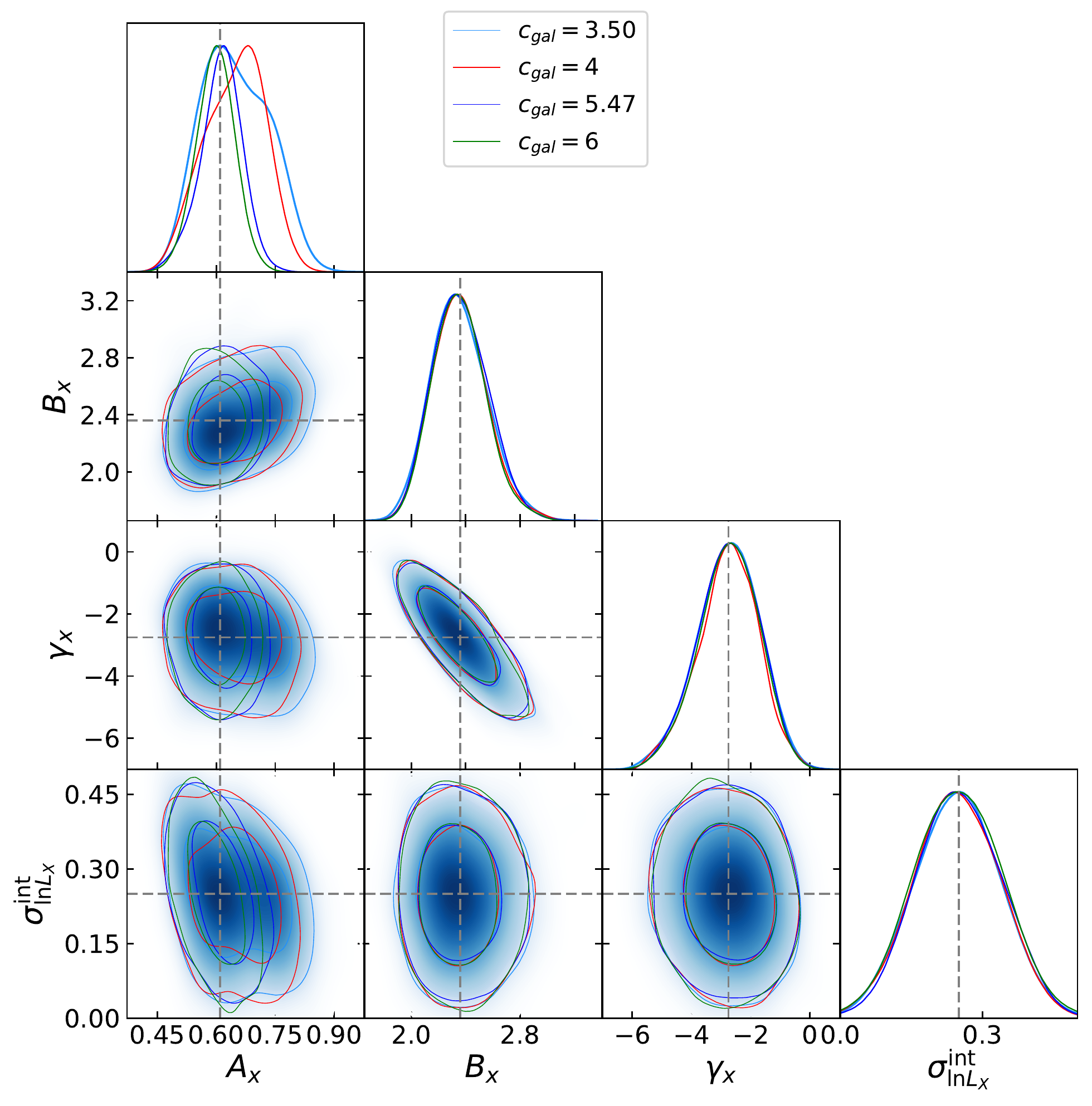} 
      }
\caption{Posterior distribution of the scaling relation parameters for varying values of $c_{\text{gal}}$. Contours show the 1$\sigma$, 2$\sigma$, and 3$\sigma$ confidence regions. Grey dashed lines correspond to the results obtained by fixing the concentration of cluster galaxies to $c_{\text{gal}}=5.37$. }
\label{fig:corner_plot_cgal}
\end{figure}

In Section~\ref{sec:number_density} we described how we account for the number density profile $\nu (r)$ of the tracer population. We assume that the number density profile of the red sequence population is described by an NFW model, with a  concentration for cluster galaxies of $c_{\text{gal}} = 5.37^{+0.27}_{-0.24}$ \citet{2017Hennig}. We test the impact of this assumption on our results, performing our analysis on a range of concentrations. In Figure~\ref{fig:corner_plot_cgal} we show that the constraints of the scaling relation parameters are not very sensitive to the choice of the concentration parameter. Grey dashed lines correspond to the results obtained by fixing $c_{\text{gal}}$ to the above mentioned value from \citep{2017Hennig}.

\bibliographystyle{mn2e}
\bibliography{literature}

\end{document}